\documentclass[a4paper,prd, 11pt]{article}
\pdfoutput=1 

\usepackage{jheppub} 

\usepackage{commath}
\usepackage{amsthm,amssymb,amsmath,mathrsfs}
\usepackage{mathtools}
\usepackage{empheq}
\usepackage{tabu} 
\usepackage[most]{tcolorbox}

\DeclarePairedDelimiterX\braket[2]{\langle}{\rangle}{#1 \delimsize\vert #2}
\graphicspath{ {images/} }

\def\be {\begin{equation}}
\def\ee {\end{equation}}
\def\bea {\begin{eqnarray}}
\def\eea {\end{eqnarray}}
\def\bc {\begin{center}}
\def\ec {\end{center}}
\def\bfg {\begin{figure}}
\def\efg {\end{figure}}
\def\bi {\begin{itemize}}
\def\ei {\end{itemize}}

\def\le {\left}
\def\ri {\right}
\def\p {\partial}

\def\vs {\vspace}
\def\hs {\hspace}

\def\c  {\cdot}
\def\a  {\alpha}
\def\b  {\beta}
\def\g  {\gamma}
\def\d  {\delta}
\def\D  {\Delta}
\def\e  {\eta}
\def\de {\partial_\eta}
\def\m  {\mu}
\def\n  {\nu}

\def\w {\omega}
\def\dw{\p_\omega}

\def\s {\sigma}
\def\D{\mathrm{D}}

\title{A local action for fermionic unconstrained higher spin gauge fields in (A)dS space
}

\author
{Mojtaba Najafizadeh}

\affiliation[]
{School of Physics\\
Institute for Research in Fundamental Sciences (IPM)\\
P.O.Box 19395-5531, Tehran, Iran}

\emailAdd{mnajafizadeh@ipm.ir}

\abstract{The local free action principle for bosonic unconstrained higher spin gauge fields in $d$-dimensional (A)dS$_d$ spacetime has been already established by Segal\,. However, later on, Schuster and Toro, in 4-dimensional flat spacetime, extracted such an action from their bosonic continuous spin gauge theory, when the continuous spin parameter vanishes\,. On the other hand, similar to the Schuster-Toro's action, we found the fermionic continuous spin gauge theory in 4-dimensional flat spacetime\,. Thus it is noteworthy to take out its fermionic higher spin formulation in a limit, and generalize it to $d$-dimensional (A)dS$_d$ spacetime\,. Therefore, in this paper, we will present a similar action principle \`a la Segal for fermions in $d$-dimensional (A)dS$_d$ spacetime\,. Moreover, at the level of equations of motion, we will demonstrate how the Fronsdal and the Fang-Fronsdal equations in $d$-dimensional (A)dS$_d$ spacetime are related, respectively, to the Euler-Lagrange equations of the bosonic and fermionic higher spin actions mentioned above.}


\preprint{IPM/P-2018/039}

\begin{document}

\maketitle



\section{Introduction}

From the massless limit of the Singh-Hagen formulation \cite{Singh:1974qz,Singh:1974rc}, the local constrained Lagrangian formulation describing free bosonic (fermionic) massless higher spin fields was established by Fronsdal (Fang and Fronsdal) in flat and Anti-de-Sitter (AdS) spacetimes \cite{Fronsdal:1978rb, Fronsdal:1978vb} (\cite{F 2, Fang:1979hq}\,) in metric-like approach (see \cite{Rahman:2015pzl} for a recent review)\footnote{~For frame-like approach see the references of \cite{Vasiliev:1980as}\,-\,\cite{Aragone:1983sz} in Minkowski space, and \cite{Vasiliev:1986td}\,-\,\cite{Vasiliev:2001wa} in AdS space (see also \cite{Skvortsov:2010nh} for massless mixed-symmetry fermionic fields)\,.}. 
The constrained Lagrangian formulations include some conditions on the gauge fields and parameters, however along with them there are unconstrained formulations comprising no conditions on the gauge fields and parameters, in both the BRST and geometric approaches, in Minkowski and (A)dS spacetimes (see e.g. \cite{Ouvry:1986dv}\,-\,\cite{Reshetnyak:2018fvd})\,\footnote{~We notice that in particular the formulation proposed in \cite{Campoleoni:2012th} does not appear to be fully unconstrained (the gauge parameter is still subject to a transversality condition), however, one can parameterize the gauge symmetry of the corresponding Maxwell-like Lagrangians in terms of fully unconstrained gauge parameters so as to obtain a fully unconstrained system \cite{Francia:2013sca}\,(We thank Dario Francia for comments and pointing out this issue)\,.}. 

An unconstrained Lagrangian formulation should be useful to make links between the higher spin theories and the BRST form of the string theory \cite{Francia:2002pt, Sagnotti:2003qa}, as well as it is thought that the unconstrained Lagrangian formulation might be helpful for studying a possible Lagrangian formulation for the Vasiliev equations, describing interacting higher spin fields \cite{Vasiliev:1990en}\,-\,\cite{Vasiliev:2003ev}\,. We note that one of the main open problems in the classical field theory is constructing a Lagrangian formalism to describe interacting higher spin fields\,\footnote{~For example, it was shown no massless higher spin field can be consistently coupled to gravity in flat space \cite{Porrati:2008rm}\,.} (see e.g. \cite{Vasiliev:2004qz}\,-\,\cite{Didenko:2014dwa} for reviews), and for that we shall concentrate on unconstrained Lagrangian formulations\,. 

Among all unconstrained Lagrangian formulations, there is a simple model for the massless free bosonic higher spin fields in $d$-dimentional (A)dS$_d$ spacetime, suggested by Segal in 2001, in terms of unconstrained gauge fields and parameters \cite{Segal:2001qq}\,. 
The merit of the Segal formulation is its simplicity, the fact that it does not introduce any auxiliary fields, and particularly its relation to the bosonic continuous spin gauge theory in a limit, explained below\,.
In fact, Segal constructed a generating formulation describing an infinite sum of the actions for massless integer-spin fields, $s=0,1,\dots,\infty$\,, such that every spin enters only one time. The obtained action includes an integral localized on the ``constraint surface'' $p^2-1=0$ in the cotangent bundle of AdS space\,. It was shown that the action can be thought as a decomposition of an infinite sum over all Fronsdal actions in (A)dS$_d$ spacetime\,. It should be again emphasized that the Segal's unconstrained formulation is devoid of any auxiliary fields, differently from
other formulations (e.g. there exist indeed two auxiliary fields in the unconstrained Lagrangian formulation presented in \cite{Francia:2007qt})\footnote{~We thank again Dario Francia for highlighting this point\,.}\,.

Later on, in 2014, Schuster and Toro in their researches on constructing a free action principle for bosonic continuous spin particle (CSP) presented an unconstrained formulation for bosonic higher spin gauge fields in flat spacetime \cite{ST PRD}\,. This action principle was obtained from a limit of the CSP theory when the continuous spin parameter (characterized usually by $\m$) vanishes ($\m=0$), and is a reformulation of the Segal's 4-dimensional action with vanishing cosmological constant ($\Lambda=0$)\,. They also recovered this action as a sum of the Fronsdal actions in another approach\,. Indeed, they integrated out the auxiliary space dependence of the action to reproduce the Schwinger-Fronsdal tensor actions, using a field decomposition and partially gauge fixing\,. 

Afterwards, in 2015, we found a similar unconstrained Lagrangian formulation for fermionic CSP in four-dimensional flat spacetime \cite{BNS}\,. Consequently, we motivated to construct a Segal-like formulation for fermionic higher spin gauge fields in $d$-dimensional (A)dS$_d$ spacetime (explained below), which has not been already discussed in the literature\,.

These above mentioned actions \cite{Segal:2001qq}\,-\,\cite{BNS} as well as the present work, can be placed in the following table  
\begin{center}
	\begin{tabular}{ c | c | c  }
		\textcolor{red}{for bosons} & $\Lambda=0$ & $\Lambda \neq 0$ \\ 
		\hline 
		$\m=0$ & Segal action \cite{Segal:2001qq} & Segal action \cite{Segal:2001qq} \\  
		$\m \neq 0$ & CSP action\,\, \cite{ST PRD}  & $?$  \\
		\hline 
		\textcolor{red}{for fermions} &  &  \\ 
		\hline 
		$\m=0$ & present work & present work \\  
		$\m \neq 0$ & CSP action\,\, \cite{BNS}  & $?$ 
	\end{tabular}
\end{center}
by understanding the fact that, \`a la Segal, the bosonic and fermionic CSP actions in (A)dS spacetime ($\m , \Lambda \neq 0$) have not been discovered so far\,. However we note that, \`a la Fronsdal, the bosonic \cite{Metsaev: B CSP} and fermionic \cite{Metsaev: F CSP} CSP actions in $d$-dimensional (A)dS$_d$ spacetime have been already established by Metsaev\,.

Therefore, to a great extent, the present work was motivated by the previous studies on the bosonic \cite{ST PRD} and fermionic \cite{BNS} continuous spin gauge theories, which are formulated in an unconstrained formalism\,.
Indeed, we realize the fact that, when the continuous spin parameter goes to zero\,($\m=0$)\,, the bosonic \cite{ST PRD} and fermionic \cite{BNS} CSP actions reduce, respectively, to the bosonic \cite{Segal:2001qq} and fermionic (present work) higher spin gauge theories in 4-dimensional flat spacetime\,.

\vspace{.1cm}
In the present work, we obtain a local and covariant action principle for free fermionic higher spin gauge fields in $d$-dimensional (A)dS$_d$ spacetime as
\vs{.1cm}
\begin{tcolorbox}[ams align, colback=white!98!black]
	{S}&=\int d^dx\,d^d\e~e~\overline{\mathbf\Psi}(x,\e)~\delta'(\e^2+1)\le(\g\cdot\e+i\,\ri)\bigg[\,\g\cdot \D - (\g\cdot\e -i\,) \,(\p_\e\cdot \D)  \label{F Action}\\ 
	&~~~~~~~~~ \quad\quad\quad\quad\quad\quad +\, \frac{i\,\sqrt{\Lambda}}{2}\, \Big(2 N_\e +d-4+(\g \c\e)(\g\c\p_\e)-3\,i\,(\g\c\de)\Big)\,\bigg]\mathbf\Psi(x,\e)\nonumber\,,
\end{tcolorbox}
\hspace{-.9cm}
where $\e^a$ is a $d$-dimensional auxiliary Lorentz vector localized to the unit hyperboloid $\e^2=-1$\,, $\delta'$ is the derivative of the Dirac delta function with respect to its argument $\delta'(a)=\frac{d}{da}\,\delta(a)$, and $\g^a$ are gamma matrices in $d$ dimensions\,. In addition, we define the Dirac adjoint as $\overline{\mathbf\Psi}:={\mathbf\Psi}^\dagger\g^0$\,, $\de^a:={\p}/{\p {\e_a}}$\,, $N_\e:=\e\c\de$\,, and $e:= \hbox{det}\, e^a_\m$\,, where $e^a_\m$ stands for vielbein of (A)dS$_d$ space\,. We also denote $\D_a:=e^\m_a\,\D_\m$ where $\D_\m$ stands for fermionic Lorentz covariant derivative, defined in the Appendix \ref{conv.}, and\,
\be 
\Lambda = \left\{
\begin{array}{ll}
	+\,1, & \hbox{for dS space;} \\
	~~\,0, & \hbox{for flat space;} \\
	-\,1, & \hbox{for AdS space,} \label{Lambda}
\end{array}
\right.
\ee
such that the (A)dS radius is set to be one\,. The gauge field $\mathbf\Psi$ is unconstrained and introduces as the generating function 
\be 
\mathbf\Psi(x,\e)=\sum_{n=0}^{\infty}\,\frac{1}{n!}~\e^{a_1} \dots \e^{a_n}~\Psi_{a_1 \dots a_n}(x)\,, \label{Psi eta}
\ee 
where $\Psi_{a_1 \dots a_n}(x)$ are totally symmetric spinor-tensor fields of all half-integer spin fields $s=n+\frac{1}{2}$\,, in such a way that the spinor index is left implicit\,. Note that in the infinite tower of spins \eqref{Psi eta}, every spin state interns only once, and the spin states are not mixed under the Lorentz boost\,\footnote{~We note that in the context of the continuous spin gauge theory, the gauge field has a form of the one in \eqref{Psi eta}, but with the difference that the spin states are mixed under the Lorentz boost\,.}.

The action \eqref{F Action} is invariant under the gauge transformations
\begin{align} 
\delta_{\xi_1} \mathbf\Psi(x,\e) &= \bigg[\,(\g\c\D)\,(\g\c\e-i\,)-(\e^2+1)\,(\de\c\D) \label {gauge T1} \\ 
&\quad\quad\quad\quad\quad+\,\frac{i\,\sqrt{\Lambda}}{2}\,\le[\,2 (\g\c\e) + (\g\c\e+i\,)^2 \,(\g\c\de)-(\g\c\e+i\,)\,(2 N_\e + d\,)  \, \ri]\bigg]\,\boldsymbol\xi_1(x,\e)\,, \nonumber \\[10pt]
 \delta_{\xi_2} \mathbf\Psi(x,\e)&=(\e^2+1)(\g\c\e+i\,)\,\boldsymbol\xi_2(x,\e)\,,\label{gauge T2}
\end{align} 
where $\boldsymbol\xi_1$ and $\boldsymbol\xi_2$ are the unconstrained gauge transformation parameters\,. Varying the action \eqref{F Action} with respect to the gauge field $\mathbf{\Psi}$ yields the following equation of motion
\begin{align}
&\delta'(\e^2+1)\le(\g\cdot\e+i\,\ri)\bigg[\,\g\cdot \D - (\g\cdot\e -i\,) \,(\p_\e\cdot \D) \label{F EOM} \\ 
&~~~~\quad\quad~ \quad\quad\quad\quad\quad\quad\quad\quad +\, \frac{i\,\sqrt{\Lambda}}{2}\, \Big(2 N_\e +d-4+(\g \c\e)(\g\c\p_\e)-3\,i\,(\g\c\de)\Big)\,\bigg]\mathbf\Psi(x,\e)=0\,. \nonumber
\end{align}
We will show later how this equation of motion can be related to the Fang-Fronsdal equation\,.

The structure of this paper is as follows. Since we aim to find a Segal-like formulation for fermions, we will first focus on the Segal action, by presenting the method in Sec.\,\ref{segal-like 1}\,. Then we will extend the method to the fermionic case and find a fermionic action in Sec.\,\ref{section 4}\,, which is the main point of this paper\,. 
Strictly speaking, in Sec.\,\ref{segal-like 1} we explain how to find the origin of the Segal action, by starting from the Fronsdal one\,. For this purpose, we will present two steps; the first step is constructing the Fronsdal-like formulation (subsec. \ref{111}), and the second one is building an unconstrained system by Fourier transforming of the Fronsdal-like formulation (subsec. \ref{222})\,. We have realized that pursuing these two steps will lead to the Segal action \cite{Segal:2001qq}\,. The reason why constructing the Fronsdal-like formulation (step one), leading to the Segal's action, is necessary clarifies in \ref{referee}\,.
In Sec.\,\ref{section 4} we will elaborate how to achieve the presented action in \eqref{F Action}\,. Therefore, similar to the bosonic case, we will first start from the Fang-Fronsdal formulation and, by presenting these two steps (Subsecs. \ref{333} and \ref{444}), will then arrive at the fermionic action \eqref{F Action}\,. 
Conclusions and further directions will be presented in Sec.\,\ref{s8}\,. We present our conventions in the Appendix \ref{conv.}\,. Useful relations are given in the Appendix \ref{Useful}\,. In order to be self-contained, a short discussion on the Segal action \cite{Segal:2001qq} will be presented in the Appendix \ref{Segal section}\,. 

\section{The Segal action} \label{segal-like 1}

In this section, we aim to construct the Segal action \cite{Segal:2001qq} from the Fronsdal one \cite{Fronsdal:1978vb}\,. In fact, at the level of equations of motion, we will make a relationship between the Fronsdal equation and the obtained Euler-Lagrange equation of the Segal action. Therefore, we first briefly review the Fronsdal formulation in $d$-dimensional (A)dS$_d$ spacetime, in which the gauge fields and parameters obey traceless constraints. However, at the end, by solving the trace conditions, we will arrive at the Segal formulation, which is devoid of any constraints on the gauge fields and parameters\,.  
To this end, we will present two steps\,. In the first step, we apply a field redefinition and construct a so-called Fronsdal-like formulation in such a way that the gauge field and parameter satisfy shifted traceless conditions. In the second step, we perform a Fourier transformation on the auxiliary space, solve the shifted traceless conditions using distributions, and construct an unconstrained formulation for the equation of motion\,. We will find the obtained equation of motion is nothing but the Euler-Lagrange equation of the Segal action\,. Finally, we explain why without the Fronsdal-like formulation we shall not be able to arrive at a correct action describing massless bosonic higher spin fields\,

\subsection{Fronsdal formulation} \label{11-}

The action describing an arbitrary massless spin-$s$ field in the Minkowski \cite{Fronsdal:1978rb} and (A)dS \cite{Fronsdal:1978vb} spacetimes were first found by Fronsdal in the metric-like approach\,. In $d$-dimensional (A)dS$_d$ spacetime, the free action for a given spin $s$ reads\,
\be
{I}_s\,=\,\frac{s!}{\,2\,}~\int \, d^{d}x ~e~ {\varphi}_s (x,\dw) ~\le[1-\,\tfrac{1}{4}\, \w^2\,(\dw\cdot\dw) \ri] \mathcal{F}_{(s)} ~{\varphi}_s (x,\w) ~\bigg|_{\w=0}\,,  \label{action}
\ee
where\,\footnote{~The operator $\mathcal{F}_s$ in \eqref{F operator} was first found in \cite{Metsaev:1999ui}\,. See also the presented formulation in \cite{Metsaev:2008fs}\,.} 
\bea
\mathcal{F}_{(s)} &=& -~ \Box_{(A)dS} \,+\, (\w \cdot \nabla)(\dw \cdot \nabla) \,-\, \frac{1}{2}\, (\w \cdot \nabla)^2\, (\dw \cdot \dw) \label{F operator} \\
&& -\, \Lambda \,\le[\, s^2 + s(d-6) - 2(d-3) + \w^2\,(\dw\cdot\dw)\,\ri]\,, \nonumber
\eea
is the Fronsdal operator\,. The operator $\Box_{(A)dS}$ denotes the D'Alembert operator of (A)dS space, and $\nabla_a:=e^\m_a ~\nabla_\m$ while $\nabla_\m$ stands for the Lorentz covariant derivative (see the Appendix \ref{conv.})\,.
The gauge field ${\varphi}_s (x,\w)$ in \eqref{action} is double-traceless 
\be
\left( \p_\omega \cdot \p_\omega \right)^2 \, \varphi_s (x,\omega) = 0\,,\quad \quad\quad
\varphi_s (x,\omega)=\frac{1}{s!} \, \, \omega^{a_{1}}   \ldots \omega^{a_{s}} \, \,
\varphi_{a_{1}\ldots a_{s}}(x)\,,   \label{generating func. 1}
\ee
describing a totally symmetric double-traceless tensor field $\varphi_{a_{1}\ldots a_{s}}(x)$ of any integer spin $s$\,, where $\omega^a$ is a $d$-dimensional auxiliary vector and $\dw^a:={\p}/{\p {\w_a}}$\,. The action \eqref{action} is invariant under the gauge transformation
\be
\delta_\epsilon \,\varphi_s (x,\omega)\,=\,\left(\omega \cdot \nabla \right)\, \epsilon_s (x,\omega)\,,   \label{Bosonic GT}
\ee
where $\epsilon_s$ is the gauge transformation parameter subject to the traceless condition 
\be
\left( \p_\omega \cdot \p_\omega \right) \, \epsilon_s (x,\omega) = 0\,, \quad\quad\quad
\epsilon_s(x,\omega)=\tfrac{1}{(s-1)!} \, \, \omega^{a_{1}}   \ldots \omega^{a_{s-1}} \, \,
\epsilon_{a_{1}\ldots a_{s-1}}(x)  \,,
\label{generating func. 2}
\ee
for the rank-$(s-1)$ symmetric and traceless gauge parameter $\epsilon_{a_{1}\ldots a_{s-1}}(x)$\,. We note that the generating functions in \eqref{generating func. 1} and \eqref{generating func. 2} satisfy, respectively, the following homogeneity conditions:
\be
\left( N_\w-s\, \right) \, \varphi_s (x,\omega) = 0\,,\quad \quad\quad\quad\quad
\left( N_\w -s+1\, \right) \, \epsilon_s (x,\omega) = 0\,, \label{homogeneity}
\ee 
where $N_\w:=\w\cdot\dw$\,. We end up here the constrained Fronsdal formulation and present a Fronsdal-like system in next subsection\,.

\subsection{Fronsdal-like formulation} \label{111}

By the Fronsdal-like formulation for a massless higher spin field we mean a formulation, in which the gauge field and parameter are redefined objects by the operators $\mathbf{P}_\Phi$ and $\mathbf{P}_\varepsilon$  
\bea 
&&\Phi_s (x,\omega)=\mathbf{P}_\Phi~\varphi_s (x,\omega)\,, \quad \quad\quad
~\mathbf{P}_\Phi= \sum_{n=0}^{\infty}~\omega^{\,2n}~ \frac{1}{~2^{\,2n} ~ n!~ (N_\w+\tfrac{d}{2}-1)_n~} 
\label{P_phi text}\,,
\\
&&\varepsilon_s (x,\omega)=\mathbf{P}_\varepsilon~\epsilon_s (x,\omega)\,, \quad \quad\quad~~\,
~\mathbf{P}_\varepsilon= \sum_{n=0}^{\infty}~\omega^{\,2n}~ \frac{1}{~2^{\,2n} ~ n!~ (N_\w+\tfrac{d}{2}\,)_n~} 
\label{P_epsi text}\,,
\eea
so that the new gauge field $\Phi_s$ and parameter $\varepsilon_s$\, satisfy, respectively, the shifted traceless conditions
\be 
\left( \p_\omega \cdot \p_\omega -1\,\right)^2  \Phi_s (x,\omega) = 0\,, \quad \quad\quad\quad\quad
\left( \p_\omega \cdot \p_\omega -1\,\right)  \varepsilon_s (x,\omega) = 0\,. \label{shifted trace}
\ee 
Indeed, the operators $\mathbf{P}_\Phi$ and $\mathbf{P}_\varepsilon$ play a role to convert the shifted traceless conditions \eqref{shifted trace} to the traceless ones \eqref{generating func. 1}, \eqref{generating func. 2}\,; and $(a)_n$ in the denominators denotes the rising Pochhammer symbol \eqref{Pochhammer}\,(for specific details of the calculation, see the appendices of \cite{Najafizadeh:2017tin})\,.

Considering a similar generating function, as before, for the redefined gauge field $\Phi_s$ and parameter $\varepsilon_s$\,, we can find the ``Fronsdal-like equation''\,\footnote{~Note that the Fronsdal-like formulation, describing a single bosonic continuous spin particle (CSP), was first discussed in \cite{BM}\,. That formulation satisfies similar conditions as \eqref{shifted trace} for the bosonic CSP gauge field and the parameter. In this sense, we called here our formulation the ``Fronsdal-like formulation'', however we note that it actually describes the bosonic higher spin gauge theory\,.}  
\bea 
&&\bigg[-\, \Box_{(A)dS}  + (\w \cdot \nabla)(\dw \cdot \nabla) - \frac{1}{2}\, (\w \cdot \nabla)^2\, (\dw \cdot \dw -1\,) \label{eom N 1}\\
&& ~\, -\, \mathrm{\Lambda}  \le( s^2 + s(d-6) - 2(d-3) +\w^2\,(\dw\cdot\dw)-2\,\w^2\, \ri) 
\bigg] \Phi_s(x,\omega) = 0\,,\nonumber
\eea
which is invariant under the gauge transformation
\be
\delta_\varepsilon \,\Phi_s (x,\omega)\,=\,\left(\omega \cdot \nabla \right)\, \varepsilon_s (x,\omega)\,.  \label{Bosonic GT-like}
\ee
To illustrate that the obtained Fronsdal-like equation \eqref{eom N 1} will reproduce the Fronsdal equation, one can first use the homogeneity condition $(N_\w -s)\,\Phi_s=0$ to rewrite the equation \eqref{eom N 1} in terms of $N_\w$\,. Then, by plugging \eqref{P_phi text} into \eqref{eom N 1}, and applying the relations \eqref{1}-\eqref{4}, it is straightforward to demonstrate that the Fronsdal-like equation \eqref{eom N 1} will precisely recover the Fronsdal equation, $\mathcal{F}_{(s)}~ \varphi_s (x,\omega) = 0$\,, up to terms of order $\mathcal{O}(\w^4)$ vanishing due to the double-traceless condition $\varphi(x,\dw)(\w^2 )^2=0$ at the level of the action \eqref{action}\,. 

The relations \eqref{shifted trace} - \eqref{Bosonic GT-like} are formulated for an arbitrary integer spin-$s$ field $\Phi_s (x,\omega)$\,. However we might be interested in a formulation where its gauge field
\be 
\Phi (x,\omega)=\sum_{s=0}^{\infty}\,\Phi_s (x,\omega)=\sum_{s=0}^{\infty}~\frac{1}{s!} \, \, \omega^{a_{1}}   \ldots \omega^{a_{s}} \, \,
\Phi_{a_{1}\ldots a_{s}}(x)\,, \label{gauge field omega}
\ee 
decomposes into infinite tower of all integer spins (s = 0,1,2, \ldots, $\infty$), in which every spin state interns only once and the spin states of the symmetric tensor fields $\Phi_{a_{1}\ldots a_{s}}(x)$ are not mixed under the Lorentz boost\,\footnote{~We note that this is in contrast to the bosonic continuous spin gauge field, where the gauge field has a similar decomposition as \eqref{gauge field omega}, but actually the spin states of the symmetric tensor fields $\Phi_{a_{1}\ldots a_{s}}(x)$ are mixed under the Lorentz boost such that the degree of mixing is determined by a continuous spin parameter\,.}.
Therefore, to construct such a formulation, we first use the homogeneity condition of the gauge field \eqref{homogeneity} into \eqref{eom N 1}, and then take into account an infinite sum over all integer spins in the relations \eqref{shifted trace} - \eqref{Bosonic GT-like}\,. The obtained result gives us the traceless conditions
\be 
\left( \p_\omega \cdot \p_\omega -1\,\right)^2  \Phi (x,\omega) = 0\,, \quad \quad\quad\quad\quad
\left( \p_\omega \cdot \p_\omega -1\,\right)  \varepsilon (x,\omega) = 0\,, \label{shifted trace 1}
\ee
the Fronsdal-like equation 
\bea 
&&\bigg[-\, \Box_{(A)dS}  + (\w \cdot \nabla)(\dw \cdot \nabla) - \frac{1}{2}\, (\w \cdot \nabla)^2\, (\dw \cdot \dw -1\,) \label{eom N 1 2}\\
&& ~\, -\, \mathrm{\Lambda}  \le( N_\w^2 + N_\w(d-6) - 2(d-3) +\w^2\,(\dw\cdot\dw)-2\,\w^2\, \ri) 
\bigg] \Phi(x,\omega) = 0\,,\nonumber
\eea
and the gauge transformation
\be
\delta_\varepsilon \,\Phi (x,\omega)\,=\,\left(\omega \cdot \nabla \right)\, \varepsilon (x,\omega)\,, \label{Bosonic GT-like 1}
\ee
in terms of the gauge field $\Phi$ \eqref{gauge field omega} and the gauge parameter $\varepsilon$\,\footnote{~The gauge parameter $\varepsilon$ has a similar decomposition as the gauge field $\Phi$, if one substitute $\Phi$ with $\varepsilon$, and $s$ with $s-1$ into \eqref{gauge field omega}\,.}.

In flat spacetime, it would be useful for our future purpose to take into account the equation of motion \eqref{eom N 1 2} in the momentum space and then perform a suitable gauge fixing (see the procedure applied in \cite{BM})\,. Consequently, in terms of the gauge-invariant distribution $\mathbf{\Phi}(p,\w)=\delta(p\c\w)\,{\Phi}(p,\w)$, the massless bosonic higher spin equations become:
\begin{align}
p^2\,\mathbf{\Phi}(p,\w)&=0 \,, \quad\quad\quad\quad\quad~\,\, (p\c\w)\,\mathbf{\Phi}(p,\w)=0\,, \nonumber\\
(p\c\dw)\,\mathbf{\Phi}(p,\w)&=0\,, \quad\quad\quad (\dw\c\dw-1)\,\mathbf{\Phi}(p,\w)=0\,.\label{wigner like}
\end{align}
We notice that these equations in their Fourier-transformed auxiliary space are precisely the Wigner equations \eqref{wigner like} with a vanishing continuous spin parameter\,.

It would be useful to highlight here a difference between the Fronsdal-like formulation \ref{111}, describing massless bosonic higher spin field, and the Fronsdal-like formulation describing a single bosonic continuous spin field \cite{BM}\,. Consider the Fronsdal-like equation \eqref{eom N 1 2} in $d$-dimensional flat spacetime ($\Lambda=0$)
\begin{align}
F_1~\Phi(x,\omega)&=\le[\,-\,\Box + (\w\c\p_x)(\dw\c\p_x)-\frac{1}{2}\,(\w\c\p_x)^2(\dw\c\dw-1\,)\ri]\Phi(x,\omega)=0\,. \label{frons likww}
\end{align}
Then, using \eqref{Fron lik}, and after dropping the term proportional to $(\dw\c\dw-1)^2$ which vanishes due to the double-traceless-like condition \eqref{shifted trace 1} in the action, one can easily check  that the kinetic operator 
\begin{align}
K_1&=\le[\,1\,-\,\frac{1}{4}\,(\w^2-1\,)(\dw\c\dw-1\,)\,\ri]F_1 \nonumber
\end{align}
is Hermitian (i.e. $K_1^\dagger=K_1$) with respect to the Hermitian conjugation
\be 
(\p_x)^\dagger \equiv -\,\p_x\,, \qquad\qquad (\dw)^\dagger \equiv \w \,, \qquad\qquad (\w)^\dagger \equiv \dw\,. \label{hermit}
\ee 
Thus one can write the action as
\be 
{I}_1 = \frac{1}{2}\,\int d^d x~\Phi(x,\dw)~ K_1~\Phi(x,\omega)\bigg|_{\w=0}\,. \label{action like}
\ee 
Therefore, the Fronsdal-like equation \eqref{frons likww} of the higher spin gauge field theory can be derived from the action \eqref{action like}\,. However, by giving the Fronsdal-like equation of the continuous spin gauge theory \cite{BM} 
\begin{align}
\mathbb{F}_1~\Phi(x,\omega)&=\le[\,-\,\Box + (\w\c\p_x+i\,\m)(\dw\c\p_x)-\frac{1}{2}\,(\w\c\p_x+i\,\m)^2(\dw\c\dw-1\,)\ri]\Phi(x,\omega)=0\,, \label{f-mu}
\end{align}
where $\m$ stands for the continuous spin parameter, using \eqref{Fron lik mu}, one can simply check that the operator 
\be 
\mathbb{K}_1=\le[\,1\,-\,\frac{1}{4}\,(\w^2-1\,)(\dw\c\dw-1\,)\,\ri]\mathbb{F}_1\,
\ee
is not Hermitian, and thus the equation \eqref{f-mu} can not be directly obtained from an action principle\,. In other words, the Fronsdal-like formulation \ref{111} of the higher spin theory is a Lagrangian formulation, while the one for the CSP theory \cite{BM} is not\,. 
In order to find a Lagrangian formulation for the CSP theory, we performed the Fourier transformation on the auxiliary space $\w$, and worked in $\e$-space, as well as solved the (double-)traceless condition using the Dirac delta distributions\,. This procedure has been explained in detail for CSP in \cite{BMN}, and here, in next subsection, we will apply it for higher spin and demonstrate how it leads to the Segal action\,.

\subsection{Unconstrained formulation} \label{222}

The Segal action is an unconstrained formulation for describing the bosonic higher spin gauge field in $d$-dimensional (A)dS$_d$ spacetime \cite{Segal:2001qq}\,, which is formulated in $\e$-space\,. Thus we will write the Fronsdal-like formulation in $\e$-space, by performing the Fourier transformation on the auxiliary space $\w$ 
\be
\widetilde{\Phi} (x,\e) = \int \frac{d^d\omega}{(2\pi)^\frac{d}2}\,\exp(\,-\,i\,\eta\cdot\omega\,)\,\Phi (x,\omega)\,. \label{fourier}
\ee

In $\e$-space, the shifted double-traceless condition \eqref{shifted trace 1} becomes $(\e^2+1)^2~\widetilde{\Phi} (x,\e)=0$\,, which can be generally solved by the Dirac delta distribution
\be
\widetilde{\Phi}(x,\e)=\delta'(\e^2+1)\,{\bf\Phi} (x,\e)\,, \label{Phi tilda}
\ee 
where ${\bf\Phi} (x,\e)$ is now an arbitrary unconstrained function
%
and $\delta'(a)=\dfrac{d}{da}\,\delta(a)$\,. We then take into account the Fronsdal-like equation \eqref{eom N 1 2} in its Fourier-transformed auxiliary space, which is 
\begin{align} 
& \bigg[-\, \Box_{(A)dS}  - (\p_\e \cdot \nabla)(\e \cdot \nabla) - \frac{1}{2}\, (\p_\e \cdot \nabla)^2\, (\e^2 +1\,) \label{eom N 1 eta}\\
&\quad \quad\quad\quad~~\, -\, \mathrm{\Lambda}  \le( (N_\e+d)^2 - (N_\e+d)(d-6) - 2(d-3) +(\p_\e \cdot \p_\e)\,\e^2+2\,(\p_\e \cdot \p_\e)\, \ri) 
\bigg] \widetilde{\Phi}(x,\e) = 0\,,\nonumber
\end{align}
with $N_\e:=\e\cdot\p_\e$\, and $\p_\e^{\,a}:={\p}/{{\p\e}_a}$\,.
Afterwards, by plugging \eqref{Phi tilda} into \eqref{eom N 1 eta}\,, and applying the identities \eqref{Ident 0} and \eqref{Ident 2}, we will conveniently arrive at the equation of motion 
\bea 
\widehat{\mathcal{K}}_b~{\bf\Phi}(x,\e)&=& \delta'(\e^2+1\,)\bigg[-\, \Box_{(A)dS}  + (\e \cdot \nabla)(\p_\e \cdot \nabla) - \frac{1}{2}\,(\e^2 +1\,) (\p_\e \cdot \nabla)^2\,  \label{eom eta}\\
&& ~~\quad\quad\quad\, -\, \mathrm{\Lambda}  \le( N_\e^2 + N_\e(d-6) - 2(d-3) + \e^2\,(\p_\e \cdot \p_\e)+2\,(\p_\e \cdot \p_\e)\, \ri) 
\bigg] {\bf\Phi}(x,\e) = 0\,,\nonumber
\eea
where the operator $\widehat{\mathcal{K}}_b$ is Hermitian (i.e. $\widehat{\mathcal{K}}_b^\dagger=\widehat{\mathcal{K}}_b$) with respect to the Hermitian conjugation
\be
(\p_x )^\dag \equiv - \, \p_x \,, \quad \quad\quad\quad (\p_{\e} )^\dag \equiv - \, \p_{\e} \,,\quad \quad\quad\quad {\e}^\dag \equiv {\e} \,. \label{rond}
\ee
Therefore, the obtained equation of motion \eqref{eom eta} can be derived from the action
\bea
{I}&=& \frac{1}{2}\,{ \int  d^d x \, d^d \eta }  ~e~ {\bf\Phi}(x,\e) ~ \widehat{\mathcal{K}}_b~ {\bf\Phi}(x,\e)\,  \label{action eta}\\
&=&\frac{1}{2}\,\int d^d x\, d^d\e~e ~{\bf\Phi}(x,\e) ~\delta'(\e^2+1)\, \bigg[-\,\Box_{_{(A)dS}} + \le(\e \cdot \nabla \ri)\le(\p_\e \cdot \nabla\ri)
- \frac{1}{2}\,(\e^2 +1\,) \le(\p_\e \cdot \nabla \ri)^2  \nonumber\\
\mkern-0mu&&~~~\quad\quad\quad\quad\quad\quad\quad\quad~- \mathrm{\Lambda}\, \le( N_\e^2 + N_\e (d-6 ) -2(d-3) + \e^2\,(\p_\e \cdot \p_\e) +2\,(\p_\e \cdot \p_\e) \, \ri)  \bigg] ~{\bf\Phi}(x,\e)\,. \nonumber
\eea  
This action is precisely the Segal action presented in \cite{Segal:2001qq}\,, describing the bosonic higher spin gauge field in $d$-dimensional (A)dS$_d$ spacetime (see the Appendix \ref{Segal section} for more detail on the Segal action). Note that, as Segal mentioned, we find the action \eqref{action eta} is equal to a sum of the Fronsdal actions \eqref{action} up to some coefficients
\be 
{I}=\sum_{s=0}^{\infty}\,\a_s~{I}_s\,. \label{I-s}
\ee 
It is notable that in 4-dimensional flat space $\Lambda=0$\,, the authors of \cite{ST PRD} also proposed the action \eqref{action eta} for presenting their massless bosonic higher spin formulation\,. They directly solved the integral over $\e$-space in the action \eqref{action eta} and illustrated the outcome \eqref{I-s} in another fashion\,. 

With a similar procedure as what was done in this subsection to obtain the Segal action, we can find the invariance of the action \eqref{action eta} under the gauge transformations (see the presented method in \cite{BMN} for CSP)
\bea 
&& \delta_{\varepsilon} \Phi (x,\e)\,=\left[    \,\e \cdot \nabla  -  \tfrac{1}{2}\, (\e^2+1\,)   \left(\p_\e \cdot \nabla     \right) \,\right] \varepsilon (x,\e)\,, \\[10pt]
&& \delta_{\chi} \Phi (x,\e)\,= (\,\e^2 + 1\, )^{\,2} \, \chi(x,\e)\,,
\eea 
where the gauge parameters $\varepsilon$ and $\chi$ are two unconstrained arbitrary functions\,.   

We note that, in four dimensions, the Segal action \eqref{action eta}  with vanishing cosmological constant ($\Lambda=0$) 
\bea
{I}&=&\frac{1}{2}\,\int d^4 x\, d^4\e ~{\bf{\Phi}} ~\delta'(\e^2+1)\, \bigg[-\,\Box_x + \le(\e \cdot \p_x \ri)\le(\p_\e \cdot \p_x\ri)
- \frac{1}{2}\,(\e^2 +1\,) \le(\p_\e \cdot \p_x \ri)^2   \bigg] {\bf\Phi}\,, \label{segal takht} 
\eea  
is precisely the bosonic CSP action \cite{ST PRD} with vanishing continuous spin parameter ($\m=0$)\,.



\subsection[Why Fronsdal-like formulation?]{Why Fronsdal-like formulation?\,\footnote{~We thank the referee for asking this question which has improved the paper\,.}}\label{referee}

In this subsection, we aim to explain why constructing the Fronsdal-like formulation \ref{111} was necessary to reach to a proper unconstrained system \ref{222}\,. Indeed, we shall demonstrate that starting from the Fronsdal formulation, and applying the method in \ref{222} (Fourier transforming and solving the traces), can not describe massless bosonic higher spin fields\,. For simplicity, we work in $4$-dimensional flat spacetime to show this fact in a simple way\,.


Let us first consider the Fronsdal action \cite{Fronsdal:1978rb} as
\be
I_0=\frac{1}{2}\,\int d^4 x~\Phi(x,\dw)~K_0 ~\Phi(x,\w)~\bigg|_{\w=0} \label{K-0}
\ee 
where $K_0$ is the kinetic operator given by \eqref{Fron lik 11}\,, and the gauge field $\Phi$ is given by the generating function \eqref{gauge field omega}\,, which is double-traceless
\be 
(\dw\c\dw)^2\,\Phi(x,\w)=0\,. \label{gheyd}
\ee
Due to this double-tracelessness, obviously, the kinetic operator $K_0$ is Hermitian (i.e. $K_0^\dagger=K_0$) with respect to \eqref{hermit}\,. Varying the action \eqref{K-0} with respect to the gauge field $\Phi$ yields the Fronsdal equation 
\begin{align}
{F}_{0}~\Phi(x,\omega)=\le[\,-\,\Box + (\w\c\p_x)(\dw\c\p_x)-\frac{1}{2}\,(\w\c\p_x)^2(\dw\c\dw)\ri]\Phi(x,\omega)&=0\,.\label{F-0}
\end{align}
In order to omit the constraint \eqref{gheyd} in the system and build an unconstrained formalism, one can perform the Fourier transformation \eqref{fourier}, and write the equations \eqref{gheyd}, \eqref{F-0} in $\e$-space 
\begin{align}
(\e^2\,)^2\,\widetilde{\Phi}(x,\e)&=0\,,\label{gheyd 1} \\[5pt]
{\widetilde{F}}_{0}~\widetilde{\Phi}(x,\e)=\le[\,-\,\Box - (\de\c\p_x)(\e\c\p_x)-\frac{1}{2}\,(\de\c\p_x)^2(\e^2\,)\ri]\widetilde{\Phi}(x,\e)&=0\,.\label{F-0 1}
\end{align}
Then, using the Dirac delta function’s property $a^2\d'(a)=0$\,, one can solve the constraint \eqref{gheyd 1} generally by 
\be 
\widetilde{\Phi}(x,\e) = \d'(\e^2)\,{\Phi}(x,\e)\,, \label{gheyd 2}
\ee 
where ${\Phi}(x,\e)$ is now an arbitrary unconstrained function
\be 
{\Phi}(x,\e)=\sum_{s=0}^{\infty}\,\frac{1}{s!}~\e^{\m_1} \dots \e^{\m_s}~\Phi_{\m_1 \dots \m_s}(x)\,. \label{gener}
\ee
Plugging \eqref{gheyd 2} into \eqref{F-0 1}, we will arrive at the equation of motion
\begin{align}
{\widehat{F}}_{0}~{\Phi}(x,\e)&=\d'(\e^2)\le[\,-\,\Box + (\e\c\p_x)(\de\c\p_x)-\frac{1}{2}\,(\e^2)(\de\c\p_x)^2\ri]{\Phi}(x,\e)=0\,,\label{F-0 1 2}
\end{align}
where the operator ${\widehat{F}}_{0}$ is Hermitian with respect to the Hermitian conjugation \eqref{rond}\,. This enable us to write the free action as
\begin{align} 
\mathcal{A}&=\frac{1}{2}\,\int d^4 x \int d^4 \e~\,\d'(\e^2)~\Phi(x,\e)\le[\,-\,\Box + (\e\c\p_x)(\de\c\p_x)-\frac{1}{2}\,(\e^2)(\de\c\p_x)^2\ri]{\Phi}(x,\e)\,. \label{action eta 2}
\end{align}
We note that this bosonic action \eqref{action eta 2}, and its fermionic analogue, were first presented in \cite{BMN}\,. This is the model that one expects to describe massless bosonic higher spin fields\,. To show this, one can follow the procedure in \cite{ST PRD}, and integrate out the auxiliary space dependence of the action \eqref{action eta 2} to see whether it reproduce the Schwinger-Fronsdal tensor actions or not\,. For this purpose, we can use a more general form of $\e$-integrals over the Euclidean space \cite{BMN} (for technical details, see the appendices in \cite{ST PRD}, \cite{BMN}\,)\footnote{~Ref. \cite{ST PRD} has discussed the case with $\s=1$, while Ref. \cite{BMN} has taken into account a generic $\s$\,.}   
\be 
\int d^4 \e~ \d'(\e^2 + \s)\,F(\e) = \le[J_0\le(\sqrt{\s\,(\de\c\de)\,}\,\ri)F(\e)\ri]\Bigg|_{\e=0} \label{limit sigma}
\ee 
where $F(\e)$ is any smooth function, and $J_0$ is the Bessel function of the first kind, of index 0\,. In the limit $\s\rightarrow 0$\,, which is the case we are studying here, it is clear that \eqref{limit sigma} becomes  
\be 
\int d^4 \e~ \d'(\e^2)\,F(\e) = F(\e)~\Big|_{\e=0}\,. \label{limit sigma 1}
\ee
Therefore, by applying \eqref{gener} and \eqref{limit sigma 1} in \eqref{action eta 2}, one can easily find that the action \eqref{action eta 2} reduces to 
\begin{align}
\mathcal{A}&=\frac{1}{2}\,\int d^4 x\, \le(\Phi(x,\e)\le[\,-\,\Box + (\e\c\p_x)(\de\c\p_x)-\tfrac{1}{2}\,(\e^2)(\de\c\p_x)^2\ri]{\Phi}(x,\e)\ri)\bigg|_{\e=0} \\
&= -\, \frac{1}{2}\,\int d^4 x~ \Phi(x)\,\,\Box\,\,\Phi(x)\,,
\end{align}
which is the Klein-Gordon action, describing the massless scalar field\,. Thus, we conclude that the action \eqref{action eta 2}, localized on $\d'(\e^2)$, does not describe the massless higher spin fields\,\footnote{~It is noteworthy to address here such a problem in context of the continuous spin gauge theory, where the first bosonic CSP action \cite{Schuster:2013pta} (localized on $\d'(\e^2)$) was problematic, and then, its modified version (localized on $\d'(\e^2+1)$) was appeared \cite{ST PRD}\,.}. However, in 4-dimensional flat spacetime, one can start the above procedure by the Fronsdal-like equation \eqref{eom N 1 2}, instead of the Fronsdal one, and obtains the Segal action \eqref{segal takht}\,.  
That action is localized on $\d'(\e^2+1)$\,, rather than $\d'(\e^2)$\,, for which one should consider \eqref{limit sigma} with $\s=1$ to integrate out the $\e$-space dependence of the action\,. This case has been studied in detail in \cite{ST PRD}, illustrating that the Segal action \eqref{segal takht} reproduces precisely the Schwinger-Fronsdal tensor actions\,. 

Another observation demonstrating that the bosonic action \eqref{action eta 2} is problematic was addressed in \cite{BMN}\,. Indeed, we investigated that the action \eqref{action eta 2} does not reproduce the correct current-current exchange for bosonic higher spins\,.

A similar discussion as above can be applied for the fermionic case (next section), and observe that constructing the Fang-Fronsdal-like formulation \ref{333} is required, however we do not pursue such a discussion in this manuscript\,.

\section{The fermionic action
} \label{section 4}

In this section, we will construct the fermionic action presented in \eqref{F Action}\,, by following the similar steps which led to the Segal action\,. For this purpose, at the level of equations of motion, we will demonstrate a relationship between the Fang-Fronsdal equation \cite{Fang:1979hq} and the obtained equation of motion in \eqref{F EOM}\,. Following the previous section for the bosonic case, we first review the Fang-Fronsdal formulation. Then we present two steps\,. In the first step, we use a field redefinition and construct the Fang-Fronsdal-like formulation\,. In the next step, by performing a Fourier transformation and solving the gamma trace conditions, we will arrive at the equation of motion \eqref{F EOM}\,, which can be directly derived from the fermionic action \eqref{F Action}\,.

\subsection{Fang-Fronsdal formulation} \label{22-}

The action describing an arbitrary massless half-integer spin field $s=n+\tfrac{1}{2}$ in $d$-dimensional (A)dS$_d$ spacetime was first proposed by Fang and Fronsdal \cite{Fang:1979hq} in metric-like approach\,\footnote{~To clarify the equivalence of metric- and frame-like formulations of higher spin fermions see e.g. \cite{Rahman:2017cxk}\,.}\,. The free action is given by 
\bea
\hspace{-.5cm}{S}_n=\int \, d^{d}x~e ~ {\overline\psi}_n (x,\dw) \le[\,1-\frac{1}{2}\,(\g\cdot\w)(\g\cdot\dw)-\frac{1}{4}\, \w^2\,(\dw\cdot\dw) \ri] \mathcal{F}_{(n)}~{\psi}_n (x,\w) \,\bigg|_{\w=0}\,,  \label{action 2}
\eea
where
\be 
\mathcal{F}_{(n)}= i\,\g\cdot \D\,-\,i\,(\w\cdot\D)(\g\cdot\dw)\, - \,\frac{1}{2}\, \sqrt{\Lambda}~\Big[\,2n+d-4+(\g\cdot\w)(\g\cdot\dw)\,\Big]
\ee 
is the Fang-Fronsdal operator (see the Appendix \ref{conv.} for conventions)\,. The action \eqref{action 2} is invariant under the gauge transformation
\be 
\delta_\zeta\, \psi_n(x,\w)=\le(\w\c\D\,+\,\frac{i\,\sqrt{\Lambda}}{2}~\g\c\w\ri)\zeta_n(x,\w)\,,
\ee 
where the spinor gauge field $\psi_n$ and parameter $\zeta_n$, using an auxiliary vector $\w^a$, are introduced as the generating functions
\begin{align} 
\psi_n (x,\omega)&=\frac{1}{n!} \, \, \omega^{a_{1}}   \ldots \omega^{a_{n}} \, \,
\psi_{a_{1}\ldots a_{n}}(x)\,, \label{psi gf}
\\[5pt] 
\zeta_n (x,\omega)&=\frac{1}{(n-1)!} \, \, \omega^{a_{1}}   \ldots \omega^{a_{n-1}} \, \,
\zeta_{a_{1}\ldots a_{n-1}}(x)  \,, \label{xi gf}
\end{align}
obeying the gamma traceless conditions
\be 
(\g\c\dw)^3\,\psi_n(x,\w)=0\,, \quad\quad\quad\quad\quad (\g\c\dw)\,\zeta_n(x,\w)=0\,,
\ee 
and the homogeneity ones
\be 
\le(N_\w-n\ri){\psi}_n (x,\w)=0\,, \label{homog cn.} \quad\quad\quad\quad\quad 
\le(N_\w-n+1\ri){\zeta}_n (x,\w)=0\,. 
\ee 
Note the spinor indices are left implicit, and $\psi_{a_{1}\ldots a_{n}}$ in \eqref{psi gf} denotes totally symmetric spinor-tensor field of half-integer spin, while $\zeta_{a_{1}\ldots a_{n-1}}$ in \eqref{xi gf} stands for the relevant spinor gauge parameter\,. We also note, using the homogeneity condition on the spinor gauge field \eqref{homog cn.}, one can be shown that the action \eqref{action 2} is precisely equivalent to the Metsaev action \cite{Metsaev:2006zy}, in the limit of massless fields\,.

\subsection{Fang-Fronsdal-like formulation} \label{333}

Similar to the bosonic case, we introduce the Fang-Fronsdal-like formulation in terms of the redefined spinor gauge field and parameter
\bea 
\mkern-50mu&&\Psi_n (x,\omega)=\mathbf{P}_\Psi~\psi_n (x,\omega)\,, \quad 
\mathbf{P}_\Psi =\sum_{k=0}^{\infty}\le[(\gamma \cdot \omega)^{2k} + 2k(\gamma \cdot \omega)^{2k-1}\ri]
\frac{1}{2^{2k}k!(N_\w+\tfrac{d}{2}-1)_k} 
\label{P say},
\\
\mkern-50mu&&~\xi_n (x,\omega)=\mathbf{P}_\xi~\zeta_n (x,\omega)\,, \quad~~ 
\mathbf{P}_\xi= \sum_{k=0}^{\infty}\le[(\gamma \cdot \omega)^{2k} + 2k(\gamma \cdot \omega)^{2k-1}\ri]
\frac{1}{2^{2k}k!(N_\w+\tfrac{d}{2}\,)_k} 
\label{P kay}\,,
\eea
such that the new spinor gauge field $\Psi_n$ and parameter $\xi_n$ satisfy, respectively, the shifted gamma traceless conditions
\be 
(\g\c\dw-1\,)(\dw\c\dw-1)\,\Psi_n(x,\w)=0\,, \label{gamma tracless} \quad\quad\quad\quad\quad (\g\c\dw-1\,)\,\xi_n(x,\w)=0\,.
\ee
The ``Fang-Fronsdal-like equation'' can be then find as\,\footnote{~Note again that the Fang-Fronsdal-like formulation, describing a single fermionic continuous spin particle (CSP), was first discussed in \cite{BM}\,. That formulation satisfies similar conditions as \eqref{gamma tracless} for the fermionic CSP gauge field and parameter. In this sense, we called here our formulation the ``Fang-Fronsdal-like formulation'', however it actually describes the fermionic higher spin gauge theory\,.} 
\be 
\bigg[\,i\,\g\cdot\D\,-\,i\,(\w\cdot\D)(\g\cdot\dw-1\,) -\,\frac{1}{2}\, \sqrt{\Lambda}~\Big[\,2n+d-4 + (\g\cdot\w) (\g\cdot\dw) - 3\, (\g\c\w)\Big]\,\bigg]{\Psi}_n (x,\w)=0\,, \label{fang-frons}
\ee 
which is invariant under the gauge transformation
\be 
\delta_\xi\, \Psi_n(x,\w)=\le(\w\c\D\,+\,\frac{i\,\sqrt{\Lambda}}{2}~\g\c\w\ri)\xi_n(x,\w)\,.
\ee
To demonstrate the obtained Fang-Fronsdal-like equation \eqref{fang-frons} is equivalent to the Fang-Fronsdal one $\mathcal{F}_{(n)}~{\psi}_n=0$\,, we can first use the homogeneity condition $\le(N_\w-n\ri){\Psi}_n (x,\w)=0$ within \eqref{fang-frons}\,. Then by plugging \eqref{P say} into \eqref{fang-frons}, and applying the relations \eqref{1 1} - \eqref{4 4}, the Fang-Fronsdal equation, $\mathcal{F}_{(n)}~{\psi}_n=0$\,, will be conveniently reproduced (up to terms of order $\mathcal{O}(\w^3)$ vanishing at the level of the action, due to the triple gamma-trace condition on the spinor gauge field $\psi_n(x,\dw)\,(\g\c\w)^3=0$)\,.

If we are interested in a formulation in terms of the spinor gauge field
\be 
\Psi (x,\omega)=\sum_{n=0}^{\infty}\,\Psi_n (x,\omega)=\sum_{n=0}^{\infty}~\frac{1}{n!} \, \, \omega^{a_{1}}   \ldots \omega^{a_{n}} \, \,
\Psi_{a_{1}\ldots a_{n}}(x)\,,
\ee 
comprising an infinite tower of all half-integer spins, and consider a similar decomposition for the spinor gauge parameter $\xi$, the shifted gamma traceless conditions read 
\be 
(\g\c\dw-1\,)(\dw\c\dw-1)\,\Psi(x,\w)=0\,, \label{gamma tracless 1} \quad\quad\quad\quad\quad (\g\c\dw-1\,)\,\xi(x,\w)=0\,.
\ee
The Fang-Fronsdal-like equation, therefore, becomes 
\be
\bigg[\,i\,\g\cdot\D\,-\,i\,(\w\cdot\D)(\g\cdot\dw-1\,) -\,\frac{1}{2}\, \sqrt{\Lambda}~\Big[\,2N_\w+d-4 + (\g\cdot\w) (\g\cdot\dw) - 3\, (\g\c\w)\Big]\,\bigg]{\Psi} (x,\w)=0\,, \label{fang-frons 1}
\ee
with the gauge symmetry
\be 
\delta_\xi\, \Psi(x,\w)=\le(\w\c\D\,+\,\frac{i\,\sqrt{\Lambda}}{2}~\g\c\w\ri)\xi(x,\w)\,.
\ee

Again as the previous section, considering the flat space limit of the equation of motion \eqref{fang-frons 1} in the momentum space, one can find the massless fermionic higher spin equations
\begin{align}
(\g\c p)\,\mathbf{\Psi}(p,\w)&=0 \,, \quad\quad\quad\quad\quad~\,\, (p\c\w)\,\mathbf{\Psi}(p,\w)=0\,, \nonumber\\
(p\c\dw)\,\mathbf{\Psi}(p,\w)&=0\,, ~\,\quad\quad\quad (\g\c\dw-1)\,\mathbf{\Psi}(p,\w)=0\,, \label{Wignerlike}
\end{align}
in terms of the gauge-invariant distribution $\mathbf{\Psi}=\delta(p\c\w)\,{\Psi}$\,. We note again that these equations are the massless higher spin limit of the Wigner equations \cite{Bargmann:1948ck}\,, if we replace the fourth equation by $(\dw\c\dw-1)\,\mathbf{\Psi}(p,\w)=0$\,(for details see the explanations in \cite{BM})\,.

\subsection{Unconstrained formulation} \label{444}

In $\e$-space, the triple-gamma traceless condition \eqref{gamma tracless 1} on the spinor gauge field becomes $(\g\c\e+i\,)(\e^2+1\,)\widetilde{\Psi} (x,\e)=0$\,, which can be generally solved by  
\be 
\widetilde{\Psi} (x,\e)=\delta'(\e^2+1\,)(\g\c\e-i\,)\,{\bf\Psi} (x,\e)\,, \label{psi tild}
\ee 
where ${\bf\Psi}$ is an unconstrained arbitrary function\,. We then take into account the Fang-Fronsdal-like equation \eqref{fang-frons 1} in its Fourier transformed auxiliary space, which is
\be
\bigg[\,i\,\g\cdot\D\,+\,i\,(\de\cdot\D)(\g\cdot\e\,+\,i\,)+\,\frac{1}{2}\, \sqrt{\Lambda}~\Big[\,2N_\e+d+4 + (\g\cdot\de) (\g\cdot\e) + 3\,i\, (\g\c\de)\,\Big]\,\bigg]\widetilde{\Psi} (x,\e)=0\,. \label{fang-frons 2}
\ee
Plugging \eqref{psi tild} into \eqref{fang-frons 2}, and applying the identities \eqref{Ident 1} and \eqref{Ident 2}, we will arrive at the equation of motion 
\bea
\mathcal{\widehat{K}}_f~{\bf\Psi}(x,\e)&=&\delta'(\e^2+1)\le(\g\cdot\e+i\,\ri)\bigg[\,\g\cdot \D - (\g\cdot\e -i\,) \,(\p_\e\cdot \D) \label{F EOM 11} \\ 
&&\quad\quad\quad\quad\quad\quad\quad\quad+\,\, \frac{i\,\sqrt{\Lambda}}{2}\, \Big(2 N_\e +d-4+(\g \c\e)(\g\c\p_\e)-3\,i\,(\g\c\de)\Big)\,\bigg]{\bf\Psi}(x,\e)=0\,. \nonumber
\eea 
This equation of motion is precisely the one in \eqref{F EOM}\,, which obtained from the action \eqref{F Action}\,. Therefore, at the level of the equation of motions, we illustrated how the Fang-Fronsdal equation, $\mathcal{F}_{(n)}~{\psi}_n=0$, can be related to the Euler-Lagrange equation of \eqref{F EOM}\,. Practically, to find the action \eqref{F Action}, we indeed found that $
\widehat{\mathcal{K}}_f^\dag = \gamma^0 \, \widehat{\mathcal{K}}_f \,  \gamma^0
$ with respect to the Hermitian conjugation \eqref{rond}\,. Using this fact, we were be able to write the fermionic action \eqref{F Action} \`a la Segal as
\bea 
{S}  &=&    { \int  d^d x \, d^d \eta }  \,\,e~ \overline{\bf\Psi}(x,\e) ~ \widehat{\mathcal{K}}_f~ {\bf\Psi}(x,\e)\,\\
&=& \int d^dx\,d^d\e~e~\overline{\mathbf\Psi}(x,\e)~\delta'(\e^2+1)\le(\g\cdot\e+i\,\ri)\bigg[\,\g\cdot \D - (\g\cdot\e -i\,) \,(\p_\e\cdot \D)  \nonumber\\ 
&&~~~~~ \quad\quad\quad\quad\quad\quad\quad\quad\quad\quad\quad\quad +\, \frac{i\,\sqrt{\Lambda}}{2}\, \Big(2 N_\e +d-4+(\g \c\e)(\g\c\p_\e)-3\,i\,(\g\c\de)\Big)\,\bigg]\mathbf\Psi(x,\e)\nonumber\,.
\eea
We note that, similar to the bosonic case, we find the action \eqref{F Action} is equal to a sum of the Fang-Fronsdal actions \eqref{action 2} with some coefficients
\be 
S=\sum_{n=0}^{\infty}\,\b_n~S_n\,. \label{I-ss}
\ee  

A similar procedure can be easily done to obtain the gauge symmetries \eqref{gauge T1}, \eqref{gauge T2}\,. For instance, in order to obviously see the invariance of the action \eqref{F Action}, and consequently the equation of motion \eqref{F EOM}, under the gauge symmetry \eqref{gauge T2}, we can use the identities \eqref{Ident 1} and \eqref{Ident 2}, to simplify the equation of motion \eqref{F EOM}, after some calculations, to the following form 
\bea
&&\bigg[\,\g\cdot \D \,+\, (\p_\e\cdot \D)\,(\g\cdot\e + i\,) \label{F EOM 2} \\ 
&&~~\quad\quad\, -\, \frac{i\,\sqrt{\Lambda}}{2}\, \Big(2 N_\e +d+4+(\g \c\de)(\g\c\e)+3\,i\,(\g\c\de)\Big)\,\bigg]\delta'(\e^2+1)\le(\g\cdot\e - i\,\ri)\mathbf\Psi(x,\e)=0\,. \nonumber
\eea 
Then, at a glance, this equation would be clearly invariant under the $\boldsymbol\xi_2$ symmetry \eqref{gauge T2}, by applying the Dirac delta function's property: $a^2\,\delta'(a)=0$\,. 

In the end, similar to the bosonic case, we refer to the fermionic action \eqref{F Action} in 4-dimensional flat spacetime
\be
S=\int d^4x\,d^4\e~\overline{\mathbf\Psi}(x,\e)~\delta'(\e^2+1)\le(\g\cdot\e+i\,\ri)\Big[\,\g\cdot \p_x - (\g\cdot\e -i\,) \,(\p_\e\cdot \p_x)\,\Big]\mathbf\Psi(x,\e)\nonumber\,,
\ee
and realize that it is indeed the fermionic continuous spin action \cite{BNS} when the continuous spin parameter vanishes ($\m=0$)\,.   

\section{Conclusions and future directions}\label{s8}

In this work, we aimed to develop the Segal unconstrained Lagrangian formulation, describing free massless bosonic higher spin fields, to the fermionic case\,. Therefore, we first explained how to find the Segal action, known as an unconstrained formulation, from the Fronsdal action, known as a constrained formulation\,. This finding was obtained in two stages\,. At the first stage, we applied a field redefinition on the Fronsdal equation and built the so-called Fronsdal-like formulation, in which the gauge field and parameter were traceless-like, instead of being traceless\,. We also discovered that the Fronsdal-like formulation is necessary in order to arrive at a correct action (Segal action)\,. In the next stage, we solved the traceless-like conditions in terms of distributions to get rid of the conditions on the gauge fields and parameters\,. Then we rewrote the Fronsdal-like formulation on its Fourier-transformed auxiliary space and obtained an unconstrained formulation\,. We note that one can introduce many other types of operators, like the operator $\mathbf{P}_\Phi$ in \eqref{P_phi text}, and build many kinds of Fronsdal-like formulations\,. However these formulations could not be led to the Segal formulation, and in this sense, the operator $\mathbf{P}_\Phi$ and consequently the Fronsdal-like formulation we found here were unique (we used precisely this operator in the case of CSP theory in \cite{Najafizadeh:2017tin})\,.       

In the end, similar to the bosonic case, we presented two steps (i.e. constructing the Fang-Fronsdal-like formulation and its Fourier transforming) to find a Segal-like action principle for fermions \eqref{F Action}, describing free fermionic higher spin gauge fields in $d$-dimentional (A)dS$_d$ spacetime, which had not been previously done in the literature\,. We found the action \eqref{F Action} is invariant under the gauge symmetries \eqref{gauge T1}, \eqref{gauge T2} where both the gauge field and the parameter were unconstrained\,.

It should be emphasized that, indeed, we made a relationship between the Fronsdal equation and the Euler-Lagrange equation of \eqref{eom eta}, as well as a connection between the Fang-Fronsdal equation and the Euler-Lagrange equation of \eqref{F EOM}\,. However making a connection at the level of the actions is still an open problem; i.e. link the Segal action \eqref{action eta} to the Fronsdal action \eqref{action} (or link the fermionic action \eqref{F Action} to the Fang-Fronsdal one \eqref{action 2})\,.

In 4-dimensional flat space, the authors of \cite{ST PRD} have directly shown that solving the $\e$-dependent part of the Segal action will lead to a direct sum of all Fronsdal actions\,\footnote{~We note that the integral on the auxiliary space in the Segal action can not be solved in the Lorentzian signature (see appendices of \cite{BMN} for more detail)\,.}. However, It would be interesting to investigate and illustrate explicitly that their applied fashion will work for the fermionic action \eqref{F Action} as well\,; i.e. do the integral on the auxiliary space in the action \eqref{F Action}, and reproduce a direct sum of all Fang-Fronsdal actions\,. Extending the manner to the higher spin theories in $d$-dimensional de-Sitter and anti-de-Sitter backgrounds would be attractive too\,.

The fermionic action presented here, together with the bosonic action of Segal, can be applied to construct supersymmetric higher spin theories in this approach, which its formulation presumably seems to be simpler than other existing theories\,. Moreover, it would be interesting to generalize the bosonic and fermionic formulations, \`a la Segal, to the partially-massless, mixed-symmetry and massive fields\,. 
In addition, as we noted in the introduction, it would be interesting to investigate whether we can, \`a la Segal, generalize the CSP theory to (A)dS spacetime\,? 

\acknowledgments

We are grateful to R. Metsaev for discussions and useful comments on an earlier draft of the paper, and to X. Bekaert and J. Mourad for collaboration on continuous spin particle which motivated this paper\,. The author is also grateful to M.M. Sheikh-jabbari for support and encouragement, to H.R. Afshar and K. Hajian for discussions, to D. Francia and A. Reshetnyak for comments, and to the School of Physics at IPM, where this work was carried out, for the warm hospitality during his visit. The referees are also acknowledged for constructive questions that have improved the paper\,.

\appendix


\section{Conventions} \label{conv.}
Our conventions are as follows\,. $x^a$ and $\e^a$ (or its Fourier transformed $\w^a$) denote respectively coordinates and auxiliary coordinates in $d$-dimensional flat spacetime, where the Latin (flat) indices take values: $a=0,1,\ldots,d-1$\,. Derivatives with respect to $x^a$ and $\e^a$ are defined as: $\p_a:={\p}/{\p x^a}$\,, ${\de}_a:={\p}/{\p {\e^a}}$\,. We use the {\bf mostly minus} signature for the flat metric tensor $\e^{ab}$\,, and define the operators $N_\e:=\e\c\de$ and $N_\w:=\w\c\dw$\,. 

The {\bf{bosonic}} covariant derivative $\nabla_a$ is given by
\be 
\nabla_a:=e^\m_a~\nabla_\m\,, \quad\quad\quad \nabla_\m:={\p}/{\p x^\m}+\frac{1}{2}~\w^{ab}_\m ~ \mathrm{M}_{ab}\,,\quad\quad\quad
\mathrm{M}^{ab}:= \e^a\,{\de}^b -\, \e^b\,{\de}^a\,,
\ee 
where $e^\m_a$ is inverse vielbein of (A)dS$_d$ space$, \nabla_\m$ stands for the Lorentz covariant derivative, $\w^{ab}_\m$ is the Lorentz connection of (A)dS$_d$ space, and $\mathrm{M}^{ab}$ denotes the spin operator of the Lorentz algebra, while the Greek (curved) indices take values: $\m=0,1,\ldots,d-1$\,. The D'Alembert operator of (A)dS$_d$ space $\Box_{(A)dS}$ is defined by
\be 
\Box_{(A)dS} := \nabla^a\,\nabla_a + e^\m_a\,\w^{ab}_\m \,\nabla_b\,.
\ee 
Flat and curved indices of the covariant totally symmetric tensor fields of (A)dS$_d$ spacetime are related to each other as: $\Phi_{a_1 \dots a_s}(x)=e^{\m_1}_{a_1}\dots e^{\m_s}_{a_s}\,\Phi_{\m_1 \dots \m_s}(x)$\,.

The {\bf{fermionic}} covariant derivative $\D_a$ is given by
\be 
\D_a:=e^\m_a~\D_\m\,, \quad\quad \D_\m:={\p}/{\p x^\m}+\,\frac{1}{2}~\w^{ab}_\m  \le(\mathrm{M}_{ab}+\g_{ab}\ri)\,,\quad\quad
\g^{ab}:=\tfrac{1}{4}( \g^a\,{\g}^b -\, \g^b\,{\g}^a)\,,
\ee
where $\g^a$ are the $d$-dimensional Dirac gamma matrices satisfying the Clifford algebra $\{\,\g^a \,,\, \g^b\, \}=2\,\e^{ab}$\,, and 
\be
\le(\gamma^a\ri)^\dag = \gamma^0 \gamma^a \gamma^0\,, \quad\quad\quad  (\gamma^0 )^\dag= +\, \gamma^0\,, \quad\quad\quad (\gamma^i)^\dag= - \,\gamma^i\,, \quad   (i=1,\ldots,d-1)\,.
\ee

We choose the mostly minus signature for the metric, however, it would be useful to stress the bosonic and fermionic formulations in the mostly plus signature for the metric as well\,. To this end, e.g. for the bosonic action \eqref{action eta}, we have to apply the following replacements
\be 
{X} ~\longrightarrow -\,{X}\,,\quad\quad\quad\quad
{Y} ~\longrightarrow \,{Y}\,,
\ee 
where ${X}$ denotes: $\Lambda\,, \e^2,\, (\de\c\de)\,, (\de\c\nabla)\,, \Box_{_{(A)dS}} $\,; while ${Y}$ stands for: $N_\e\,, (\e\c\nabla)$\,. For the fermionic action \eqref{F Action}, the substitutions should be taken into account as
\be 
\mathcal{X} ~\longrightarrow i\,\mathcal{X}\,,\quad\quad\quad\quad
(\g\c\e) ~\longrightarrow -\,i\,(\g\c\e)
\,,
\quad\quad\quad\quad
\mathcal{Y} ~\longrightarrow -\,\mathcal{Y}
\,,
\ee  
where $\mathcal{X}$ denotes: $\g^a\,,\overline{\mathbf\Psi}\,,  (\g\c\de)\,, (\g\c\D)$\,; while $\mathcal{Y}$ stands for: $\Lambda\,, \e^2\,, (\de\c\D)$\,.

\section{Useful relations} \label{Useful}

The \textit{rising Pochhammer symbol} $(a)_n$ is defined as
\be
(a)_n := 
a \, (a+1)(a+2) \cdots (a+n-1) = \frac{\Gamma(a+n)}{\Gamma(a)} \,, \quad\quad n\in\mathbb N ~~\mbox{and}~~ a\in\mathbb R\,.  \label{Pochhammer}
\ee

The following useful relations can be conveniently derived (see the Appendix of \cite{Metsaev:2006zy} for more detail)
\be 
\hspace{- 1 cm}[\,\D^a \,,\, \e^2\,]=0\,, ~\quad\quad\quad\quad\quad\quad[\,\de^2 \,,\,\D^a \,]=0\,, ~\quad\quad\quad\quad\quad\quad[\,\D^a \,,\, \g\c\e\,]=0\,,
\ee 
\vs{0.0005 cm}
\be 
[\,\de^2 \,,\, \e\c\D\,]=2\,\de\c\D\,, \quad\quad[\,\g\c\de \,,\, \e\c\D\,]=\g\c\D\,, \quad\quad\{\,\g\c\D \,,\, \g\c\e\,\}=2\,\e\c\D\,,
\ee 
\vs{0.0005 cm}
\be 
\hs{.14cm}\le[\,\g\c\D \,,\, \e\c\D\,\ri]=\Lambda\le(\g\c\e\le[N_\e+\frac{d-1}{2}\ri] - \e^2(\g\c\de) \ri)
\ee 
\vs{0.005 cm}
\be 
\le[\,\de\c\D \,,\, \g\c\D\,\ri]=\Lambda\le(\le[N_\e+\frac{d-1}{2}\ri]\g\c\de - (\g\c\e)\,\de^2 \ri)
\ee 

The identities 
\begin{align}
(\p_\e \cdot \p_\e) \,\delta'(\e^2+1) &= \delta'(\e^2+1) \,(\p_\e \cdot \p_\e) + 4 \,\delta''(\e^2+1) \,(h-3) - 4\,\delta'''(\e^2+1)\,, \label{Ident 0}
\\[11pt] 
(\g\c\de)\,\delta'(\e^2+1)&=2\,(\g\c\e)\,\delta''(\e^2+1)+\delta'(\e^2+1)\,(\g\c\de) \,, \label{Ident 1}
\\[11pt]
N_\e \,\delta'(\e^2+1)&=\delta'(\e^2+1)\,(N_\e -4) - 2\,\delta''(\e^2+1)\,, \label{Ident 2}
\end{align}
can be easily obtained, where $\delta''(a)$ (or $\delta'''(a)$) is the derivative of $\delta'(a)$ (or $\delta''(a)$) with respect to its argument $a$\,.

The quantities $\dw^{\,a}$\,, $\dw^{\,2}$\,, $\w^a$ and $N_\w$ on the bosonic operator $\mathbf{P}_\Phi$\,, introduced in \eqref{P_phi text}\,, act as (for more detail, see the Appendices in \cite{Najafizadeh:2017tin}\,\footnote{~We note, in this paper, the metric signature is the mostly minus while, in \cite{Najafizadeh:2017tin}, the one is the mostly plus\,.})
\begin{align}
~~\dw^{\,a}~\mathbf{P}_\Phi&=\mathbf{P}_\Phi\bigg[\, \dw^{\,a}\,-\,\w^2~\frac{1}{(2N+d)(2N+d-2)}~\dw^{\,a}\,+\,\w^a~\frac{1}{(2N+d-2)}\,\bigg]\,,  \label{1}
\\[10pt]
\dw^{\,2}~\mathbf{P}_\Phi&=\mathbf{P}_\Phi\bigg[\,(\dw\cdot\dw)\,-\,\w^2~\frac{2}{(2N+d-2)(2N+d+2)}~(\dw\cdot\dw)
\nonumber \\
&\quad~~~~~~~+\,\frac{2N+d}{(2N+d-2)} \,-\,\w^2~\frac{2}{(2N+d)(2N+d-2)^2}+~  \mathcal{O}(\w^4) \bigg]\,, \label{2}
\\[10pt]
\w^a~\mathbf{P}_\Phi&=\mathbf{P}_\Phi\bigg[\,\w^a\,+\,\w^2\, \w^a~\frac{1}{(2N+d)(2N+d-2)}~+~  \mathcal{O}(\w^4) \bigg]\,, \label{3}
\\[10pt]
N_\w~\mathbf{P}_\Phi&=\mathbf{P}_\Phi\bigg[\,N_\w\,+\,\w^2\, \frac{1}{(2N+d-2)}~+~  \mathcal{O}(\w^4) \bigg]\,, \label{4}
\end{align}
where the terms containing $\mathcal{O}(\w^4)$ will be eliminated at the level of the action, due to the double-traceless condition on the gauge field ${\Phi}(x,\dw)\,(\w^2)^{\,2}\,=0$\,.
%
On the other hand, the quantities ${\g\cdot\D}$\,, ${\w\cdot\D}$\,, ${\gamma \cdot \dw}$ and $N_\w$ on the fermionic operator $\mathbf{P}_\Psi$\,, given by \eqref{P say}\,, act as
\begin{align}
({\gamma \cdot \D})~ \mathbf{P}_\Psi&=\mathbf{P}_\Psi\bigg[\, ({\gamma \cdot \D})\,+\,({\w\cdot\D})~\frac{2}{(2N+d-2)}\,-\,({\gamma\cdot\w})({\gamma \cdot \D})~\frac{2}{(2N+d-2)}\nonumber\\
&~~~\,\quad\quad\quad\quad\quad~ -\,({\gamma\cdot\w})({\w\cdot\D})~\frac{2}{(2N+d)(2N+d-2)} \nonumber\\
&\quad\quad\quad\quad\quad~~~~\, 
+ ~\w^2\,({\gamma \cdot \D})~\frac{2}{(2N+d)(2N+d-2)}~+~  \mathcal{O}(\w^3) ~ \bigg]\,,  \label{1 1}
\\[10pt]
({\w\cdot\D})~\mathbf{P}_\Psi&=\mathbf{P}_\Psi\bigg[\,({\w\cdot\D})
\,+\,({\gamma\cdot\w})({\w\cdot\D})~\frac{2}{(2N+d)(2N+d-2)}~+~  \mathcal{O}(\w^3)~ \bigg]\,, \label{2 2}
\\[10pt]
({\gamma \cdot \dw})~ \mathbf{P}_\Psi&=\mathbf{P}_\Psi\bigg[\, ({\gamma \cdot \dw})\,-\,
({\gamma \cdot \w})~\frac{2}{(2N+d-2)^2}  \nonumber \\
&\quad\quad\quad\quad\quad~~~\,~~+\,\w^2~\frac{1}{(2N+d)^2} ~ ({\gamma \cdot \dw})\,+\, \frac{2N+d}{(2N+d-2)} \nonumber \\
&\quad\quad\quad\quad\quad~~~~\,~-\,\,({\gamma \cdot \w}) \,\frac{2(2N+d-1)}{(2N+d)(2N+d-2)}\,({\gamma \cdot \dw})
\, +  \mathcal{O}(\w^3) \, \bigg]\,,  \label{3 3}
\\[10pt]
N_\w~\mathbf{P}_\Psi&=\mathbf{P}_\Psi\bigg[\,N_\w\,+\,(\g\c\w)\,\frac{1}{2N+d-2}\,+\,\w^2\,\frac{2N+d-1}{(2N+d)(2N+d-2)}~+   \mathcal{O}(\w^3) \bigg]\,, \label{4 4}
\end{align}
where the terms containing $\mathcal{O}(\w^3)$ will be vanished, at the level of the action, because of the triple gamma-trace condition on the spinor gauge field ${\overline{\Psi}}(x,\dw)\,(\g\cdot\w)^3=0$\,.

The operators $K_0$, $K_1$ and $\mathbb{K}_1$ are given as the following:
\begin{align}
\mkern-54mu{K}_{0}&=\le[\,1\,-\,\frac{1}{4}\,\w^2\,(\dw\c\dw)\,\ri]\le[\,-\,\Box + (\w\c\p_x)(\dw\c\p_x)-\frac{1}{2}\,(\w\c\p_x)^2(\dw\c\dw)\ri] \quad\label{Fron lik 11}\\
\mkern-54mu&=-\,\Box +(\w\c\p_x)(\dw\c\p_x)+\frac{1}{2}\,\le[\w^2\,\Box\,(\dw\c\dw)-(\w\c\p_x)^2\,(\dw\c\dw)-\w^2 \, (\dw\c\p_x)^2 \ri] \nonumber\\
\mkern-54mu&~~~\,+\frac{1}{4}\,\w^2\,(\w\c\p_x)(\dw\c\p_x)\,(\dw\c\dw)+\frac{1}{8}\,\w^2(\w\c\p_x)^2(\dw\c\dw)^2\,, \nonumber
\end{align}

\begin{align}
{K}_{1}&=\le[\,1\,-\,\frac{1}{4}\,(\w^2-1\,)(\dw\c\dw-1\,)\,\ri]\le[\,-\,\Box + (\w\c\p_x)(\dw\c\p_x)-\frac{1}{2}\,(\w\c\p_x)^2(\dw\c\dw-1\,)\ri] \label{Fron lik}\\
&=-\,\Box +(\w\c\p_x)(\dw\c\p_x)\nonumber\\
&+\frac{1}{2}\,\le[(\w\c\p_x)^2+(\dw\c\p_x)^2-(\w\c\p_x)^2\,(\dw\c\dw)-\w^2 \, (\dw\c\p_x)^2+\,\Box-\w^2-(\dw\c\dw)+\w^2\,\Box\,(\dw\c\dw) \ri] \nonumber\\
&+\frac{1}{4}\,\le[(\w\c\p_x)(\dw\c\p_x)-\w^2(\w\c\p_x)(\dw\c\p_x)-(\w\c\p_x)(\dw\c\p_x)(\dw\c\dw)+\w^2\,(\w\c\p_x)(\dw\c\p_x)\,(\dw\c\dw)\ri] \nonumber\\ &+\frac{1}{8}\,(\w\c\p_x)^2(\w^2-1)(\dw\c\dw-1)^2\,, \nonumber
\end{align}

\begin{align}
\mathbb{K}_1&=\le[\,1\,-\,\frac{1}{4}\,(\w^2-1\,)(\dw\c\dw-1\,)\,\ri]\le[\,-\,\Box + (\w\c\p_x+i\,\m)(\dw\c\p_x)-\frac{1}{2}\,(\w\c\p_x+i\,\m)^2(\dw\c\dw-1\,)\ri] \nonumber\\
&= K_1+i\,\m\le[\,\dw\c\p_x - (\w\c\p_x)(\dw\c\dw-1)\,\ri] +\,\frac{1}{2}\,\m^2\le[(\dw\c\dw-1)-\frac{1}{4}\,(\w^2-1)(\dw\c\dw-1)^2\ri]\nonumber\\
&~~~~~~~~+\,\frac{i\,\m}{4}\,\le[(\w^2 -1)(\dw\c\p_x)(\dw\c\dw-1)+(\w^2 -1)(\w\c\p_x)(\dw\c\dw-1)^2\ri] \,.\label{Fron lik mu}
\end{align}

\section{The Segal action in (A)dS}\label{Segal section}

In this Appendix, we review briefly the Segal action \cite{Segal:2001qq} in a more convenient form to compare with our results in Sec.\,\ref{segal-like 1}\,.

The invariant action of the bosonic higher spin gauge fields on AdS$_d$ spacetime, in the mostly negative signature\footnote{\,Note that in \cite{Segal:2001qq}, the action is written in the mostly plus signature for the metric\,.}, is given by \cite{Segal:2001qq}
\be
S=\frac{1}{2}\,\int d^d x\, d^dp~\sqrt{-g\,} ~\tilde{h} ~\delta'(p^2+1)\, \le\{-\,A B + 2 B A + V_{11} - \frac{1}{2}\le(p^2 +1\ri) \le(A^2 - V_{21}\ri)   \ri\} \tilde{h}\,,
\label{Segal action}
\ee
where $p^\m$ is an auxiliary $d$-dimensional vector\,, $g=\det(g_{\m\n})$\,, $\delta'(a)=\frac{d}{da}\delta(a)$ and the unconstrained gauge field $\tilde{h}$ is considered as the generating function
\be
\tilde{h}=\tilde{h}(x,p) = \sum_{s=0}^{\infty}\,\frac{1}{s!}\, p^{\m_1} \cdots p^{\m_s}~h_{\m_1 \ldots \m_s}(x)\,,
\ee
with $h_{\m_1 \ldots \m_s}(x)$ corresponding to totally symmetric tensor fields of all integer spins (one row Young tableaux) in any dimension $d$\,. The operators $A$, $B$, $V_{11}$ and $V_{21}$ in the action \eqref{Segal action} are given by\footnote{~There is a typo in the operator $V_{11}$ introduced in \cite{Segal:2001qq}. The corrected one is given here in \eqref{A}\,.}
\bea
&& A=\p_p \cdot \nabla\,, \quad\quad ~\,~ V_{11}=2\,\mathrm{\Lambda} \le( 2\,p\cdot \p_p + d - 3 \ri)\,,  \label{A}\\
&& B=~p \cdot \nabla   \,,   \quad\quad~~~  V_{21}=-\, 4\,\mathrm{\Lambda} \le(\p_p\cdot\p_p\ri)\,, \label{B}
\eea
where $\p_p^\m :=\frac{\p}{\p p_\m}$\,, $\mathrm{\Lambda}$ is the constant scalar curvature defined in \eqref{Lambda}, and $\nabla_\m$ is the bosonic ``covariant derivative''
\be
\nabla_\m = \p_\m + \mathrm{\Gamma}_{\m\n}^\a(x)\,p_\a\,\p_p^\n \,,  \quad\quad\quad\p_\m:=\frac{\p}{\p x^\m}\,,
\ee
with the Riemannian connection ${\Gamma}_{\m\n}^\a(x)$ corresponding to the metric $g_{\m\n}(x)$\,, so that
\be
\le[\,\nabla_\m\,,\,\nabla_\n\,\ri] f(x,p)= p_\a R_{\m\n~\,\b}^{~\,\,~\a}(x)\,\p_p^\b~ f(x,p) = \mathrm{\Lambda} \le(p_\m\,\p_{p\,\n} - p_\n\,\p_{p\,\m}\ri) f(x,p)\,.
\ee
Using the latter, and $[\,\p_p^\m\,,\,p_\n\,]=\delta^\m_{\,\,\n}$\,, it is straightforward to demonstrate the following commutator
\be
\le[\,A\,,\,B\,\ri] = \Box_{_{AdS}}+\mathrm{\Lambda} \le( N^2 + N(d-2) - p^2\,\p_p \cdot \p_p  \ri)\,,  \label{commut.}
\ee
where $N=p\cdot \p_p$ and
\be
\Box_{_{AdS}} = \nabla_\m \,\nabla^\m + 2 \, {\Gamma}_{\n\m}^\a\, p_\a\,\p_p^{\,\m}\,\nabla^\n\,.
\ee
Then, we can rewrite the action \eqref{Segal action} by substituting \eqref{A}, \eqref{B} and \eqref{commut.} in \eqref{Segal action} as
\bea
\mkern-70muS&=&\frac{1}{2}\,\int d^d x\, d^dp~\sqrt{-g\,} ~\tilde{h} ~\delta'(p^2+1)\, \bigg\{-\,\Box_{_{AdS}} + \le(p \cdot \nabla\ri)\le(\p_p \cdot \nabla\ri)
- \frac{1}{2}\,(\,p^2 +1) \le(\p_p \cdot \nabla\ri)^2  \nonumber\\
\mkern-30mu&&~~~~~\quad\quad\quad~\quad\quad\quad\quad~- \mathrm{\Lambda}\, \le( N^2 + N (d-6 ) -2(d-3) + p^2\,(\p_p \cdot \p_p) +2\,(\p_p \cdot \p_p)  \ri)  \bigg\} ~\tilde{h}\,. \label{Segal action 1}
\eea

%




\end{document}